\documentclass[11pt,twoside]{article}
\usepackage{asp2010}
\usepackage{graphicx}

\resetcounters

\begin{document}

\title{The binary central stars of PNe with the shortest orbital period}
\author{M. Santander-Garc\'\i a$^{1,2,3}$, P. Rodr\'iguez-Gil$^{1,2,3}$, D. Jones$^4$, R. L. M. Corradi$^{2,3}$, B. Miszalski$^5$, S. Pyrzas$^6$, and M. M. Rubio-D\'\i ez$^7$
\affil{$^1$Isaac Newton Group of Telescopes, Ap.\ de Correos 321, E-38700 Sta. Cruz de la Palma, Spain}
\affil{$^2$Instituto de Astrof\'\i sica de Canarias, E-38200 La Laguna, Tenerife, Spain}
\affil{$^3$Departamento de Astrof\'\i sica, Universidad de La Laguna, E-38205 La Laguna, Tenerife, Spain}
\affil{$^4$Jodrell Bank Centre for Astrophysics, School of Physics and Astronomy, The University of Manchester, Manchester, M13 9PL, UK}
\affil{$^5$Centre for Astrophysics Research, STRI, University of Hertfordshire, College Lane Campus, Hatfield AL10 9AB, UK}
\affil{$^6$Department of Physics, University of Warwick, Coventry, CV4 7AL, UK}
\affil{$^7$Centro de Astrobiolog\'\i a, CSIC-INTA, Ctra de Torrej\'on a Ajalvir km 4, E-28850 Torrej\'on de Ardoz, Spain}
}

\begin{abstract}
Close binarity can play a significant role in the shaping of planetary nebulae (PNe) as the system evolves through the common-envelope phase. We present the detection of two of the shortest orbital periods among PN binary central stars. These are Hen 2-428, a bipolar PN, and V458 Vul, a recent nova surrounded by a mildly bipolar planetary nebula. The properties of the central stars of these systems, of their nebulae and their possible fate are discussed.\end{abstract}

\section{Introduction}

The link between the shaping of bipolar planetary nebulae and their central stars (CSPN) is still poorly understood. The distinct theoretical approaches to explain their shaping (see the review by \citealp{balick02}) fall into two broad categories: {\it a)} rapid stellar rotation and/or magnetic fields \citep[e.g.][]{garciasegura99,blackman01a}, and {\it b)} a close interacting companion to the star (e.g. \citealt{nordhaus06}, for a review see \citealt{demarco09}). This binary hypothesis is currently gaining ground as the number of close binary systems at the cores of bipolar PNe is growing (e.g. \citeauthor{miszalski09b} 2009b).

In this context, our group found several new close-binary CSPNe during recent photometric observing runs at the 1.2-m Mercator telescope (\citealp{miszalski10}). One of the main goals of these runs was to test the hypothesis that certain morphological features of PNe such as rings and/or jets could be related to close binarity. Some of the newly detected binary CSPNe show orbital periods as short as a few hours.

In this work we discuss the extreme case of the shortest orbital periods of binary CSPNe through the examples of Hen~2--428, a bipolar PN, and V458~Vul, a CSPN which recently underwent a nova eruption and shows the shortest orbital period of this kind of object so far.

\begin{figure}[!h]
\center
\resizebox{13cm}{!}{\includegraphics{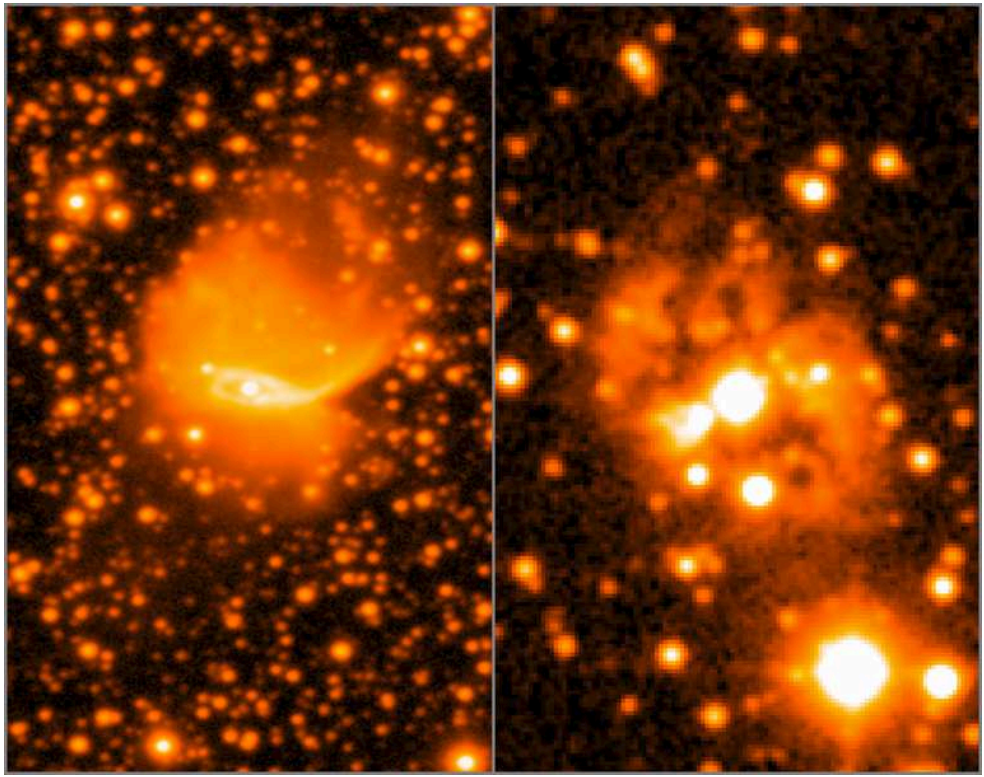}}
\caption{{\bf Left}: Close-up 2 hour-deep image of Hen~2--428 in H$\alpha$ with the INT/WFC. North is up and East is to the left. There is no apparent trace of ancient polar jets (not even at a larger scale). {\bf Right:} Close-up 36~min image of V458~Vul with the INT/WFC. North is up and East is to the left.}
\label{F1}
\end{figure}

\section{Hen~2--428}

This nebula consists of a pair of open lobes emerging from a central waist shaped as a ring inclined 68$^\circ$ to the plane of the sky. A 2-hour deep  H$\alpha$ image (see Fig.~\ref{F1}) taken with the Wide Field Camera (WFC) at the 2.5-m Isaac Newton Telescope (INT) reveals no apparent trace of ancient polar jets in the system. Hen~2--428 was studied in detail by \cite{rodriguez01}, who already hinted at the possibility of this system actually hosting a binary core. Among the properties which lead to that suggestion were the extremely low abundance of every element except He (in particular, the oxygen abundance is as low as [O/H]$\sim$7.8); the unresolved nebular core with an electron density of n$_\mathrm{e}$=10$^{3.3-6}$ cm$^{-3}$, indicative of strong mass loss or exchange phenomena taking place; and a central region with  n$_\mathrm{e}$\textgreater10$^{10}$ cm$^{-3}$ where the Ca {\sc ii} triplet arises in emission, which could be explained by the presence of an accretion disc.

The following results are preliminary and will be further analysed and discussed in \citeauthor{santander10b} (in preparation).

\subsection{Observations}

We carried out I-band photometry of Hen~2--428 with MEROPE on the Mercator telescope on the nights of August 28 and 30, 2009. Once we detected a  photometric variability as large as $\sim$0.36 mag between different short blocks, we monitored the system for a continuous, 4 hour period on September 2, 2009. Later, on May 1, 2010, we carried out a 2.25 hour spectroscopic monitoring with Intermediate dispersion Spectrograph and Imaging System (ISIS) on the 4.2-m William Herschel Telescope (WHT), with the R600B and the R316R gratings in the blue and red arms, respectively, and a slit width of 1 arcsec. The resolution was, respectively, $\sim$1.5 and $\sim$3 \AA.

\subsection{Results}

\begin{figure}[!h]
\center
\resizebox{13cm}{!}{\includegraphics{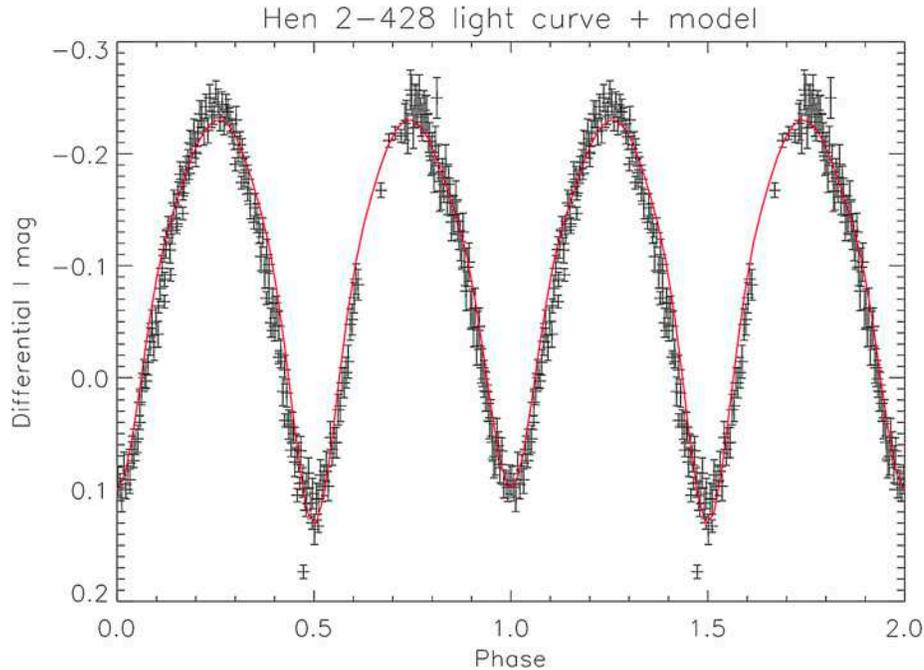}}
\caption{I-band light curve of Hen~2--428, folded on the period of 4.2 hours determined in the text. A preliminary model of the light curve is shown as a thick red line.}
\label{F2}
\end{figure}

The light curve of Hen~2--428 in the I-band, folded on the period determined below, is shown in Fig.~\ref{F2}. We performed a period analysis using Schwarzenberg-Czerny's (\citealp{Scwarzenberg96})  analysis-of-variance (AOV) method as implemented in the {\tt MIDAS/TSA} context. The AOV periodogram shows the strongest peak at  $\sim$11.379 cycles d$^{-1}$, which corresponds to a period of 0.0879 days, or 2.1 hours. However, the light curve shows a clear ellipsoidal modulation with the two minima showing almost equal depths (in a very similar case to MT~Ser, the nucleus of Abell 41, \citealt{bruch01}). This, together with the confirmation of the orbital phase given by the spectra (see below), leads us to consider a period twice as large as being the orbital period of the system. 

A sine fit to the whole data set results in P$\sim$0.17 d ($\sim$4.2 hours), and an I-band amplitude of $\sim$0.18~mag. We take the observed CSPN variability as a clear indication of binarity, the brightness modulation being the result of ellipsoidal modulation of one or both stars (i.e. the star is gravitationally distorted, close to or filling its Roche lobe and thus presents the observer at Earth a varying surface  along its relative orbital position, and therefore a varying flux). This is supported by the modelling of the light curve presented below.

The spectroscopic monitoring, carried out with the WHT/ISIS for approximately two thirds of a complete orbital period, shows an apparent double S-wave pattern which could be interpreted as two different He {\sc ii} 468.6~nm components in absorption. This would indicate that Hen~2--428  hosts in fact a double-degenerate nucleus with two white dwarves. However, the S/N ratio of the spectra are too low to either unambiguously ascertain the presence of the two absorption components or carry out precise measurements of  the Doppler shifts which could lead to mass ratio and orbital parameter constraints, for which an 8-meter class telescope will be needed.

On the other hand, a preliminary modelling of the light curve, made both with {\tt PHOEBE} (\citealt{prsa05}) and the {\tt LCURVE} code developed by Tom Marsh (\citealt{copperwheat10}) provides a fair fit to the data (see Fig.\ref{F2}). Although the lack of constraints such as the orbital velocities and the mass ratio results in a degeneracy of most of the model parameters, the following results hold for every possible set of parameters: (a) the light variations are due to ellipsoidal modulation, (b) the system is a contact binary in which both components fill their Roche lobe, (c) the T$_{\mathrm{ eff}}$ of both stars are similar within a few thousand Kelvin and (d) the inclination of the orbital plane ranges between 40$^\circ$ and 70$^\circ$, close to the 68$^\circ$-inclined equatorial ring of the nebula.

\section{V458~Vul}

V458~Vul (see Fig.~\ref{F1}) is a system which underwent a fast nova eruption on the summer of 2007. It was the subject of a work by \cite{wesson08}, in which they reported the discovery of a $\sim$ 14000 year-old, wasp-waisted PN  surrounding the nova progenitor. Their photoionisation model revealed that the ionising source must have a temperature of T$_\mathrm{eff}\sim$90000 K, a luminosity of L$_\mathrm{bol}\sim$3000 L$_{\sun}$ and a radius of R$\sim$0.23 R$_{\sun}$. These values lead, through the use of H-burning tracks, to a mass for the ionising source of M$\sim$0.58 M$_{\sun}$. On the other hand, fast nova theoretical models (e.g. \citealt{yaron05}, \citealt{prialnik95}), together with some observations (\citealp{ritter03}) point to a mass of M$>$1 M$_{\sun}$ for the nova progenitor to be able to trigger the thermonuclear runaway, which is clearly at odds with the value found via the H-burning tracks.

\subsection{Observations}

In an attempt to measure a precise orbital period we started a time-resolved spectroscopic campaign searching for the orbital signature in the radial velocities of the emission lines. The spectroscopic data were obtained with the Intermediate Dispersion Spectrograph (IDS) on the INT and ISIS on the WHT, both on La Palma. The total time on target was 39 hours, over 9 different nights in the period June 4, 2008 -- August 31, 2009. The instrumental setup covered a range of gratings and resolutions, for details see  \cite{rodriguezgil10}.

\subsection{Results}

The average optical spectrum of V458~Vul taken on 2008 June 4 (day 301 after the nova explosion) showed the flat-topped emission line profiles characteristic of this class of objects. However, the round-topped profile of the He {\sc ii} 541.2~nm line attracted our attention during a first visual inspection of the line shapes. We found the Doppler shift of this line to vary between $\sim$-200 and $\sim$200~km~s$^{-1}$ This radial velocity variation indicated that at least one of the components of the He {\sc ii} 541.2~nm emission forms in a binary system at the core of the planetary nebula. By November 2008, the He {\sc ii} 468.6~nm also showed a clear modulation, once the line thinned out (i.e. shed some nova ejecta emission).

The radial velocity data, combining the measurements from the He {\sc ii} 541.2 nm and He {\sc ii} 468.6~nm lines was subjected to the same AOV period analysis as Hen~2--428. The periodogram exhibits a narrow spike at 14.68~cycles~d$^{-1}$. A sine fit to the whole data set resulted in  P = 0.06812255$\pm$0.00000017 d (98.1 min). This makes V458~Vul the binary CSPN with the shortest orbital period known so far.

This finding presents us with an intriguing discrepancy: a nova with such a short orbital period must be an old system, of the order of Gyr, much older indeed  than the 14000 yr old PN we observe. This discrepancy is easily removed, however, by assuming the observable PN to be the second one of the system (the resulting ejecta after a second common-envelope stage). Furthermore, this is fully consistent with the measurements and modelling presented by \citealp{wesson08}: we have a double-degenerate nucleus consisting of two white dwarves, one with M$\sim$0.58 M$_{\sun}$, the present ionising source, and another one, the nova progenitor, with M$>$1 M$_{\sun}$.

Hence, the total mass of V458~Vul may well be above 1.6 M$_{\sun}$, above the critical Chandrasekhar mass, indicating that it may become a Type Ia supernova, if the white dwarf manages to accumulate mass in the presence of nova eruptions.

The aforementioned results are discussed in greater detail in \cite{rodriguezgil10}.

\section{Summary and final thoughts}

We have determined the orbital period and nature of the central stars of Hen~2--428 (0.18 days) and V458~Vul (0.068 days), the latter of which is also a SN Ia candidate. V458~Vul is the CSPN with the shortest orbital period known so far, while the period of Hen~2--428 figures among the shortest orbital periods known (after Te~11, 0.12 days, \citealt{miszalski10}; Pe~1-9 and BMP~1800-3408, 0.14 days, \citeauthor{miszalski09a} 2009a; NGC~6778,  0.15 days, \citealt{miszalski10};  SBS~1150+599A, 0.16 days, \citealt{tovmassian10} and \citealt{demarco09}).

On the other hand, both of the discussed objects appear to be double-degenerate close-binary CSPNe, a class of objects which might indeed be quite common (e.g. MT~Ser, \citealt{bruch01}; TS~01, \citealt{tovmassian10}; NGC~6026, \citeauthor{Hillwig10a} 2010a; but see also \citeauthor{Hillwig10b} 2010b), amounting up to 33\% of the sample in \citeauthor{miszalski09a} (2009a). 

Building a larger sample of this class of close-binary CSPN is important, not only to better assess their fraction in the total number of PNe, but also because they can provide us with vital information about the role of the second common-envelope stage and Roche-lobe filling (or contact binarity) in the evolution and shaping of PNe, and because some of them might be SN Ia candidates.

\acknowledgements MSG thanks Tom Marsh for the use of his code, and the SOC of the APN V conference for the invited talk.

\bibliographystyle{asp2010}
\bibliography{msantander}

\end{document}